\documentclass[12pt]{iopart} \usepackage{iopams}
\usepackage{graphicx}

\usepackage{iopams}
\usepackage{epsfig}
\def\be {\begin{equation}}
\def\ee {\end{equation}}

\def\bea {\begin{eqnarray}}
\def\eea {\end{eqnarray}}

\newcommand{\bef}{\begin{figure}}
\newcommand{\eef}{\end{figure}}
\newcommand{\ra}{\rightarrow}

\begin{document}
\title{Thermal Radiation from Au + Au Collisions at $\sqrt{s_{NN}}=200$ GeV Energy}
\author{Jan-e Alam$^a$, Jajati K. Nayak$^a$, Pradip Roy$^b$, Abhee K. Dutt-Mazumder$^b$
and Bikash Sinha$^{a,b}$
}
\medskip
\address{$a$ Variable Energy Cyclotron Centre, 1/AF Bidhan Nagar,
Kolkata 700 064, INDIA\\
$b$ Saha Institute of Nuclear Physics, 1/AF Bidhan Nagar
Kolkata 700 064, INDIA}

\begin{abstract}
The transverse momentum distribution of the direct photons measured by the
PHENIX collaboration in $Au + Au$ collisions at
$\sqrt{s_{NN}}=200$ GeV/A has been analyzed. It has been shown that
the data can be reproduced reasonably well assuming
a deconfined state of thermalized quarks and gluons 
with initial temperature more than the
transition temperature for deconfinement inferred from lattice
QCD. The value of the initial temperature depends on
the equation of state of the evolving matter. The sensitivities of 
the results on various input parameters have been studied.  
The effects of the modifications of hadronic properties at non-zero 
temperature have been discussed. 

\end{abstract}

\maketitle
\section{Introduction}

Study of photon spectra emanating from hot and dense  matter
formed in ultra-relativistic heavy ion collisions is a field of considerable
current interest. Electromagnetic probes have been proposed to be
the promising tools to characterize the initial state of 
the collisions~\cite{larry,GK,weldon90,jpr,annals}. Because of the very nature of their 
interactions
photons and dileptons suffer minimum rescattering and  
therefore, can be used as efficient tools to extract the 
initial temperature of the system.  By comparing 
the initial temperature  with the transition
temperature estimated from lattice QCD, one can infer whether Quark Gluon 
Plasma (QGP) is formed or not. 

On the experimental side substantial progress has been made in measuring
the transverse momentum spectrum of photons from nuclear collisions
at Super Proton Synchrotron (SPS)~\cite{wa98} and Relativistic Heavy 
Ion Collider 
(RHIC)~\cite{adler1}.
In contrast to the earlier results ~\cite{adler1} PHENIX collaboration
has analyzed the data by using a novel technique and reported~\cite{phenix} 
excess direct photons over the  next to leading order perturbative QCD 
(NLO pQCD) processes for $Au +Au$ collisions at $\sqrt{s_{NN}}=200$ GeV. The
corresponding theoretical analysis of  the data 
presented in Ref.~\cite{adler1} was performed in Ref.~\cite{peress}. 
The purpose of the present work
is to analyze the new experimental data and infer the initial
temperature of the system formed after the collisions. The sensitivity of the
results on various input parameters {\it e.g.} transition 
temperature, strong coupling constant, equation of states etc 
have been presented.

The paper is organized as follows. In the next section 
the photon production rates from QGP and hot hadronic gas are
discussed. The photons are produced from various reactions and decays
taking place in an expanding medium. Therefore, the static (fixed 
temperature) emission rates should be convoluted with the space
time evolution of the system. The space time evolution of the
system is outlined in section 3. Results of the calculations
are presented in section 4.  Section 5 is devoted for summary and
discussions.

\section{ Photon spectra}
Let us identify the
possible sources of ``excess'' photons above those coming
from the decays of $\pi^0$, $\eta$ mesons etc. as provided
by the data. 
Photons from the decays of $\pi^0$, $\eta$ etc. are subtracted from the data 
and hence will not be discussed here. For the transverse momentum
spectra of the photon, first we focus on the high $p_T$ domain. These are
populated by the prompt photons originating from the hard collisions of 
initial state partons in the colliding nuclei. This is believed to
be the domain where contributions can be estimated by  
pQCD.  We use  NLO predictions by Gordon and 
Vogelsang~\cite{gordon} from $pp$ collisions
and scale it up by the number of binary collisions for $Au+Au$ interactions to obtain
the prompt contributions to the direct photons.
It should be noted here that NLO prediction does not require any
intrinsic $k_T$ smearing to explain the $p-p$ data~\cite{adler}.
This effect is ignored in the analysis of $Au+Au$ data in the present work. 
The fast quarks propagating through QGP 
lose energy due to gluon radiation and  hence produce
photons with reduced energy via fragmentation processes. 
This indicates that photon production by these 
processes will be suppressed. However, the induced emission of photons
by the hard partons due to multiple scattering in the QGP will
enhance the photon radiation. It is shown in Ref.~\cite{zakharov} 
that the enhancement due to induced radiation compensates the suppression due
to jet energy loss for the $p_T$ domain considered here. 
Therefore, we ignore these mechanisms in the current analysis.
Photon production from the interaction of thermal 
gluons and non-thermal quarks was first considered in~\cite{npa1997}
within the framework of Fokker Planck equation.
Contributions  from the  hard partons
undergoing annihilation and Compton processes with the quarks and gluons 
in the thermal medium~\cite{fries} have been evaluated and 
its importance has been highlighted recently.
Discussions on these contributions are beyond the scope of the 
present work (see ~\cite{moore} for further details).
The duration of the pre-equilibrium stage will be small
because thermalization time taken here is small ($\sim 0.2$ fm) 
and hence the contributions from this stage may  be small.
Photons from the pre-equilibrium stage and hard-thermal conversion are 
 neglected here.
The fact that the current experimental data can be explained 
without these contributions indicate that theoretical
uncertainties exist at present. 

The estimation of the thermal contribution depends on the 
space-time evolution scenario
that one considers.  In case of a deconfinement phase transition, 
which seems to
be  plausible at RHIC energies (see~\cite{npa757} for a review), 
one assumes that QGP is formed initially. 
The equilibrated plasma then expands, cools,
and reverts back to hadronic matter and finally freezes out  
at a temperature $\sim$ 120 MeV. Evidently there will also be thermal radiation
from the luminous hadronic fireball which has to be evaluated properly in
order to have a reliable estimate of the initial temperature. 

The photon emission rate from QGP due to Compton ($q(\bar{q})g\rightarrow 
q(\bar{q})\gamma$) and annihilation ($q\bar{q}\rightarrow g\gamma$) processes 
was evaluated~\cite{kapusta,baier} by using Hard Thermal Loop (HTL) 
approximation~\cite{pisarski}.  Later it was shown~\cite{auranche1}
that photon production  from the reactions,
$gq\rightarrow gq\gamma$, $qq\rightarrow qq\gamma$,
$qq\bar{q}\rightarrow q\gamma$ 
and $gq\bar{q}\rightarrow g\gamma$ contribute in the same order
as annihilation and Compton processes.
However, this calculation does not
incorporate suppression due to multiple scattering during the
emission process. This point was later clarified 
in Ref.~\cite{auranche2}. The complete calculation of photon
emission rate from QGP to order $\alpha_s$ has been completed
by resuming ladder diagrams in the effective theory~\cite{arnold}. 
We use the results of  Ref.~\cite{arnold} in the present work. 
The parameterizations  of the
emission rates for various processes 
are available in Ref.~\cite{renk}. The temperature dependence of the
strong coupling constant is taken from Ref.~\cite{zantow}.

While evaluating the photons  from hadronic phase
we consider an exhaustive set of hadronic reactions and the radiative 
decay of higher resonance states
~\cite{we1,we2,we3,we4}. 
The relevant reactions and decays for photon productions are: 
(i) $\pi\,\pi\,\ra\,\rho\,\gamma$, (ii) $\pi\,\rho\,
\ra\,\pi\gamma$ (with $\pi$, $\rho$, $\omega$, $\phi$ and $a_1$ in the
intermediate state~\cite{we3}), (iii)$\pi\,\pi\,\ra\,\eta\,\gamma$ and 
(iv) $\pi\,\eta\,\ra\,\pi\,\gamma$, 
$\rho\,\ra\,\pi\,\pi\,\gamma$ and $\omega\,\ra\,\pi\,\gamma$. 
The corresponding vertices are obtained
from various phenomenological Lagrangians described in detail 
in Ref.~\cite{we1,we2,we3,we4}. 
Contributions from other decays, such as 
$K^{\ast}(892)\,\ra\, K\,\gamma$, $\phi\,\ra\,\eta\,\gamma$, 
$b_1(1235)\,\ra\,\pi\,\gamma$, $a_2(1320)\,\ra\,\pi\,\gamma$ 
and $K_1(1270)\,\ra\,\pi\,\gamma$ have been found to be 
small~\cite{haglin} for $p_T>1$ GeV. 
All the isospin
combinations for the above reactions and decays have properly been
taken into account.

Various experiments suggest that 
the spectral functions of hadrons are modified in a dense 
nuclear environment~\cite{kek,chaos,tagx,naruki,hades,clas,na60}. 
The enhancement of lepton pair yield in CERES data~\cite{ceres} below 
the $\rho$-mass can only be explained
by assuming the in-medium modifications of the $\rho$ 
meson~\cite{rw,sah,li,rapp}.
The photon spectra measured by WA98 collaboration at CERN-SPS energies
have been reproduced by assuming the reduction of hadronic masses
in the thermal bath~\cite{ja1,ja2}.
It was shown in Ref.~\cite{ja2} that the $p_T$ distribution
of photons changes significantly with a reduced mass scenario and
is almost unaffected by the broadening~\cite{ja2} of the
vector meson spectral function in the medium.
On the other hand the invariant mass distribution of the lepton pairs
is sensitive to both the reduction in vector meson  masses~\cite{sah,li} as
well as the enhanced width of the vector mesons~\cite{rw,rapp}.
Thus, by looking only at the dilepton spectra, 
it is difficult to differentiate the above scenarios. We need
to analyze both the photon and dilepton spectra simultaneously.

There is no consensus on the nature of the vector meson modification
in matter - pole shift or broadening - both experimentally~\cite{naruki,hades,
clas,na60} and theoretically~\cite{rw,harada}. The shift in hadronic
spectral function depends on the temperature and (baryonic)
chemical potential of the thermal bath created after the collisions.
The value of baryonic chemical potential for system produced 
after the collision at $\sqrt{s_{NN}}=200$ GeV  is much
smaller~\cite{npa757} compared to the other lower energy collisions~\cite{corw}.
Therefore, the extrapolation of
the nature of change observed in these low energy experiments to RHIC 
energy may suffer from various uncertainties, because 
for matter formed at RHIC collision
the net baryon number is very small (baryon - anti-baryon $\sim 0$). 
Because of these reasons and the insensitivities of
photon spectra on broadening we consider the reduction of hadronic masses
to evaluate photon productions from hadronic matter to estimate
$T_i$.  The nature of changes
in the hadronic spectral function at non-zero temperature and density is not 
known from first principle at present. Thus one has to 
rely on calculations based on various phenomenological 
models (see~\cite{annals,rw,gb} for a review).

\begin{figure}
\begin{center}
\includegraphics{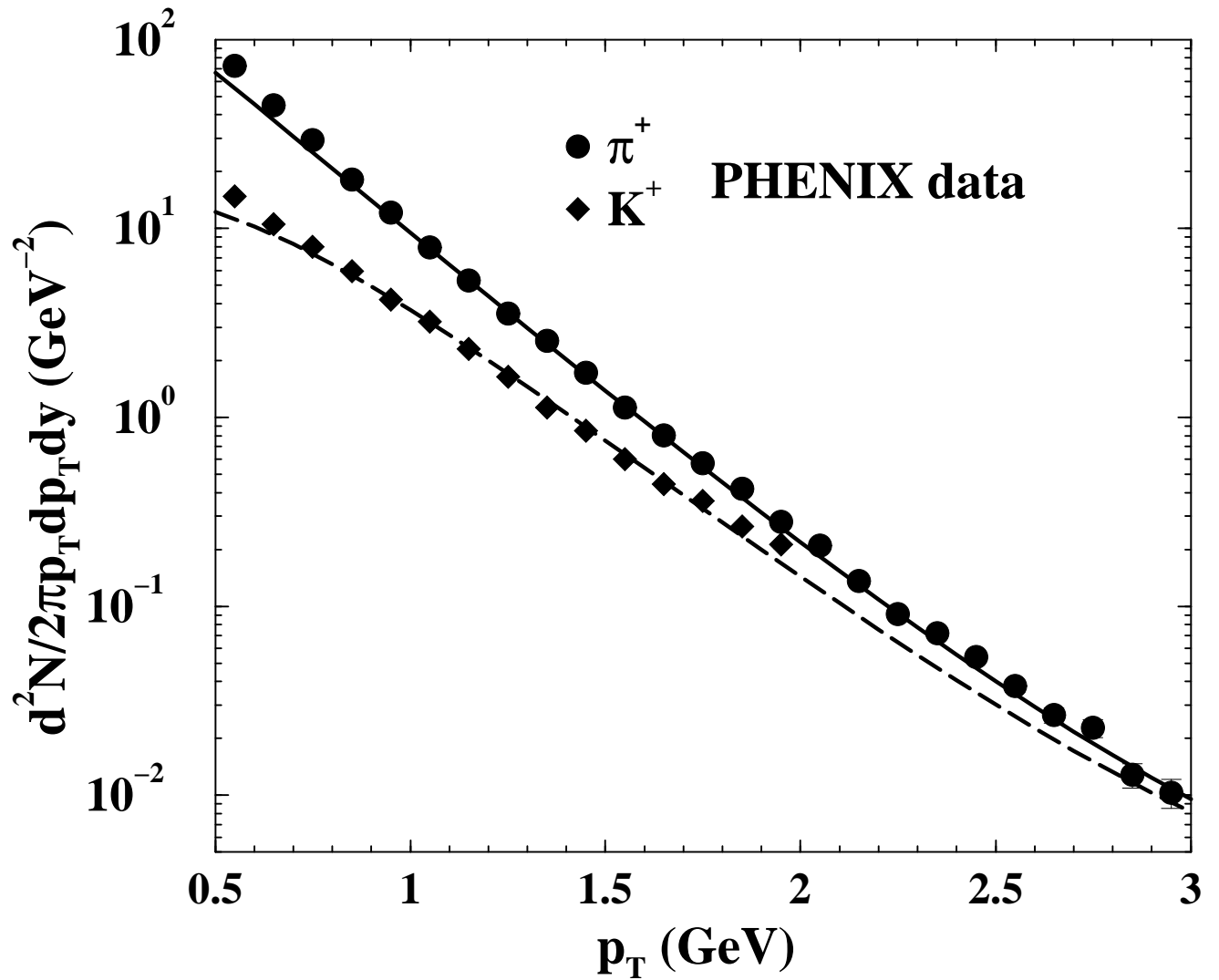}
\caption{$\pi^+$ (circle) and $K^+$ (diamond) spectra at $\sqrt{s_{NN}}=200$ GeV
measured by PHENIX Collaboration. Solid (dashed) line depicts the pion (kaon) 
spectra obtained in the hydrodynamical model. 
The data is taken from~\protect{\cite{prc69}} for Au + Au collisions at
$\sqrt{s_{NN}}=200$ GeV for $(0-5)\%$ centrality.
Type I EOS has been here with $T_i=400$ MeV, $\tau_i=0.2$ fm and $T_f=120$ MeV.
}
\label{fig1}
\end{center}
\end{figure}

In the present work we use Brown-Rho (BR) scaling scenario~\cite{brown}
(see also~\cite{brrecent})
for in-medium modifications of hadronic masses 
(except pseudoscalars).  BR scaling has been
used here to indicate how far the value of initial temperature is
affected when the reduction of the hadronic mass is incorporated
in evaluating the photon spectra. The BR scaling indicates stronger 
reduction of hadronic masses as compared to Quantum Hadrodynamics
(QHD)~\cite{vol16}. As the abundances of hadrons increase with the
reduction in their masses, photon yield is expected to increase from
hadronic phase with BR scaling.  Therefore, in this scenario
a conservative estimate of the photons from QGP phase
and hence a conservative value of the initial temperature
is obtained. In this article,
we do not discuss the details about the computational 
procedure as this has been done in Refs.~\cite{annals,we1,we2,ja1,ja2}.

\section{Space time evolution}
The ideal relativistic hydrodynamics in
(2+1)$D$~\cite{hvg} with longitudinal boost 
invariance~\cite{bjorken} and cylindrical symmetry has been used 
for the space time evolution. 
\subsection{initial condition}
In case of isentropic expansion the experimentally measured hadron multiplicity
can be related to the initial temperature and thermalization time 
by the following equation~\cite{hwa}:
\be
T_i^3(b)\tau_i=\frac{2\pi^4}{45\zeta(3)\pi\,R_A^2 4a_k}\frac{dN}{dy}(b)
\label{intemp}
\ee
where $dN/dy(b)$ is the hadron (predominantly pions) multiplicity
for a given impact parameter $b$, 
$R_A$ is the radius of the system,
$\tau_i$ is the initial thermalization time,
$\zeta(3)$ is the Riemann zeta function and
$a_k=({\pi^2}/{90})\,g_k$ is the degeneracy of the system created.
The hadron multiplicity  resulting from $Au + Au$ collisions 
is related to that from pp collision at a given 
impact parameter and collision
energy by
\be
\frac{dN}{dy}(b)=\left[(1-x)N_{part}(b)/2+xN_{coll}(b)\right]\frac{dN_{pp}}{dy}
\ee
where $x$ is the fraction of hard collisions.
$N_{part}$
is the number of participants and  $N_{coll}$ is the number of collisions 
evaluated by using Glauber model. 
$dN_{pp}^{ch}/dy= 2.5-0.25ln(s)+0.023ln^2s$,
is the multiplicity of the produced hadrons
in $pp$ collisions at centre of mass energy, $\sqrt{s}~\cite{KN}$.  
We have assumed that $25\%$ hard (i.e. $x=0.25$ )and 
$75\%$ soft collisions are responsible for initial entropy production.

We further assume that the system is formed in a thermalized phase of quarks and
gluons at  the initial thermalization
time  $\tau_i=0.2$ fm. Taking  the number of flavours, 
$N_F = 2.5$, $dN/dy\sim 1100$ and solving 
Eq.~\ref{intemp}  the value of the initial temperature ($T_i$)
is obtained as $T_i=400$ MeV. 
The initial energy density and  radial velocity profiles 
are taken as:
\be
\epsilon(\tau_i,r)=\frac{\epsilon_0}{1+e^{(r-R_A)/\delta}}
\label{enerin}
\ee
 and
\be
v(\tau_i,r) = v_0\left[1-\frac{1}{1+e^{(r-R_A)/\delta}}\right]
\label{vrin}
\ee
Sensitivities of the results on the velocity profiles  will be shown,
$\delta$ ($\sim 0.5$ fm here) is a parameter, known as the surface thickness. 
Unless mention explicitly the initial radial velocity is taken as zero.

\subsection{Equation of state (EOS)}
Apart from the initial conditions, we also need
the EOS to solve the hydrodynamic equations.
Ideally the EOS should be obtained from the first principle 
{\it i.e.} within the framework thermal QCD.
Although, these results has large uncertainties   
for temperature $\leq T_c$ because 
the pseudoscalar mass is large ~\cite{karschprivatecomm},
we have used it here to find out the conservative value of $T_i$. 
Therefore, two types of equation of state will be used here
to study the photon spectra to indicate the sensitivity 
of the results.

(I)  Bag model type EOS has been used for QGP.
For EOS of the hadronic matter all the 
resonances with mass
$<$  2.5 GeV $/c^2$ has been considered. 
The velocity of sound 
is taken as $c_s^2=1/3$ and $1/5$~\cite{bm}
for QGP and hadronic phase respectively. 
The effect of baryonic chemical potential 
is neglected here. 

(II) EOS from lattice QCD~\cite{karsch02} has also been 
used here to show the sensitivity of our results on the 
equation of state and
 to find out the conservative value of $T_i$. 

The transition temperature is taken as $ T_c\sim $ 190 MeV as 
obtained from lattice QCD ~\cite{katz,cheng} recently. However, 
the sensitivity of the results  
on $T_c$ will also be demonstrated.
The freeze-out temperature, $T_f$ has been fixed by studying 
the transverse momentum distribution of hadrons.

\section{Results}
First we use the type I EOS and initial conditions 
described above to solve the relativistic hydrodynamic equations 
for studying  the $p_T$ spectra of pions and kaons.
To reproduce the transverse momentum distribution of pions
and kaons (Fig.~\ref{fig1}) measured experimentally 
by PHENIX collaboration~\cite{prc69}, 
the required value of
$T_f\,\sim\,120$ MeV. In all the results shown below
the value of the freeze-out temperature is
fixed at 120 MeV.
The dependence of the $p_T$ distributions of hadrons
on the initial temperature is rather weak for the $p_T$ values
under consideration.  It is assumed here that the chemical 
equilibrium is maintained up to $T_f$.
A comment on the chemical freeze-out temperature is in order here. 
Depending upon the scenarios (see below) considered here,
the value of the initial temperatures are
400 MeV and 590 MeV, which are quite high. Because of the high 
initial temperatures
the photon emission from the early stage dominate over the 
contributions from the late stage of the evolution. 
This is in agreement with the results obtained in~\cite{huovinen1} 
at lower (SPS) energies. Moreover, it is also 
shown in~\cite{huovinen1} that  results from ideal hydrodynamics and 
coarse grained UrQMD are in reasonable agreement.
Therefore, photon spectra will not be affected significantly 
if chemical equilibrium is not maintained in the late stage of the
evolution. The effect of finite
baryonic chemical potential on the production rate of photon 
from an equilibrated QGP  is also small~\cite{traxler}.

\begin{figure}
\begin{center}
\includegraphics{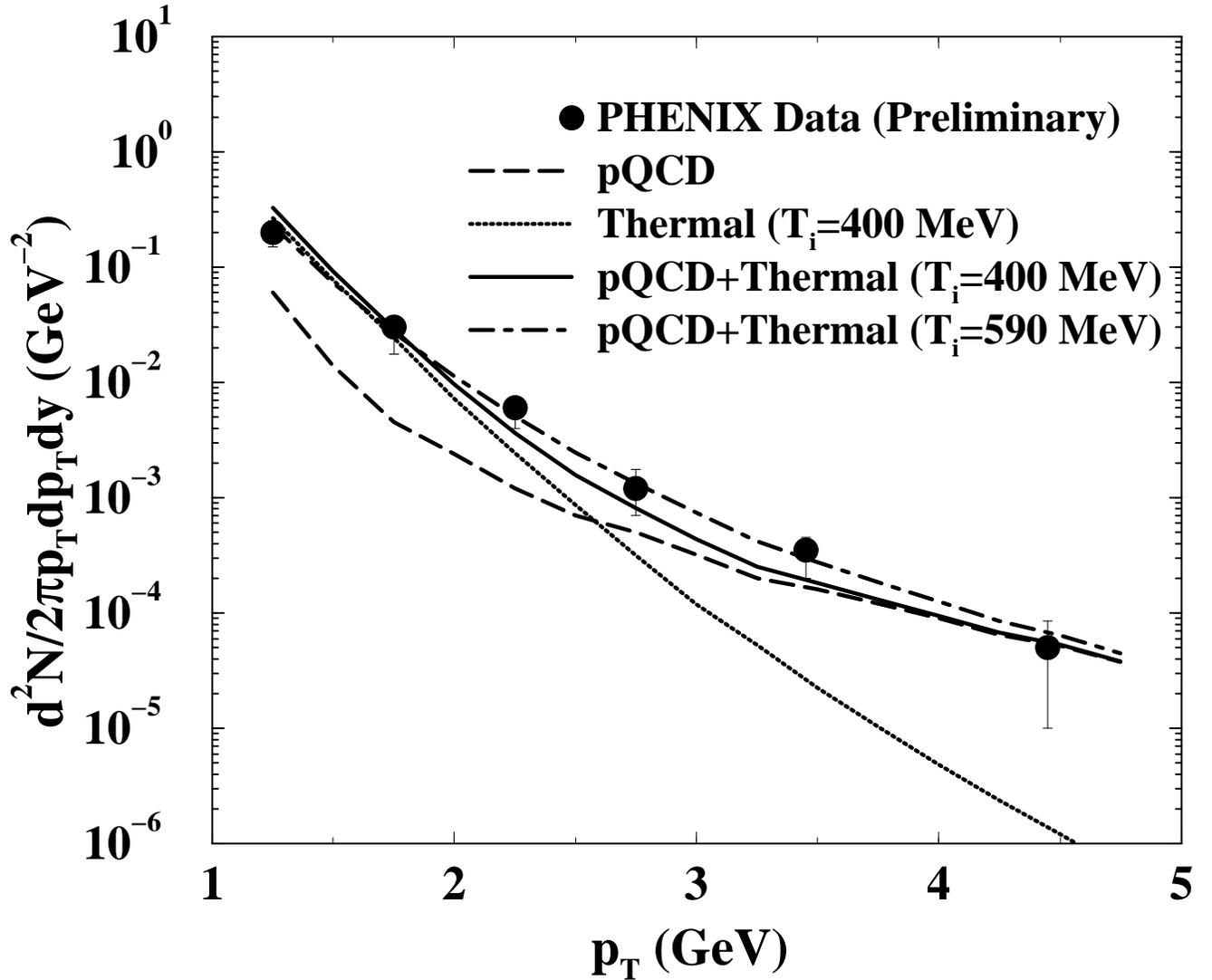}
\caption{Direct photon spectra at RHIC energies measured by 
PHENIX Collaboration for $(0-20)\%$ centrality.
Dashed line indicates hard photons from NLO pQCD 
calculations~{\protect\cite{gordon}}. 
Solid 
(dot-dashed) 
line depicts the total photon yield 
obtained from QGP initial state with $T_i$ = 400 MeV and
$\tau_i=0.2$ fm 
($T_i=590$ MeV $\tau_i=0.15$ fm). 
Type I EOS has been used to obtain the thermal contributions 
shown in this figure.
In medium effects on hadrons  are included 
(ignored) in the results shown by solid (dot-dashed) line. 
Photon production rate from QGP is taken from~{\protect\cite{arnold}}.
}

\label{fig2}
\end{center}
\end{figure}

Now we concentrate  on the transverse momentum spectra of photons.
The resulting spectra is contrasted 
with the recent PHENIX measurements of direct
photons in Fig.~\ref{fig2}. We observe that the data is
reproduced with $T_i=400$ MeV and $\tau_i=0.2$ fm 
with in-medium modification of hadrons and type I EOS.
It is found that the contributions from quark matter and
hadronic matter to the photon production are similar 
in the $p_T$ interval, $1\leq p_T$(GeV)$\leq 3$, 
the range where thermal contribution dominates. 

\begin{figure}
\begin{center}
\includegraphics{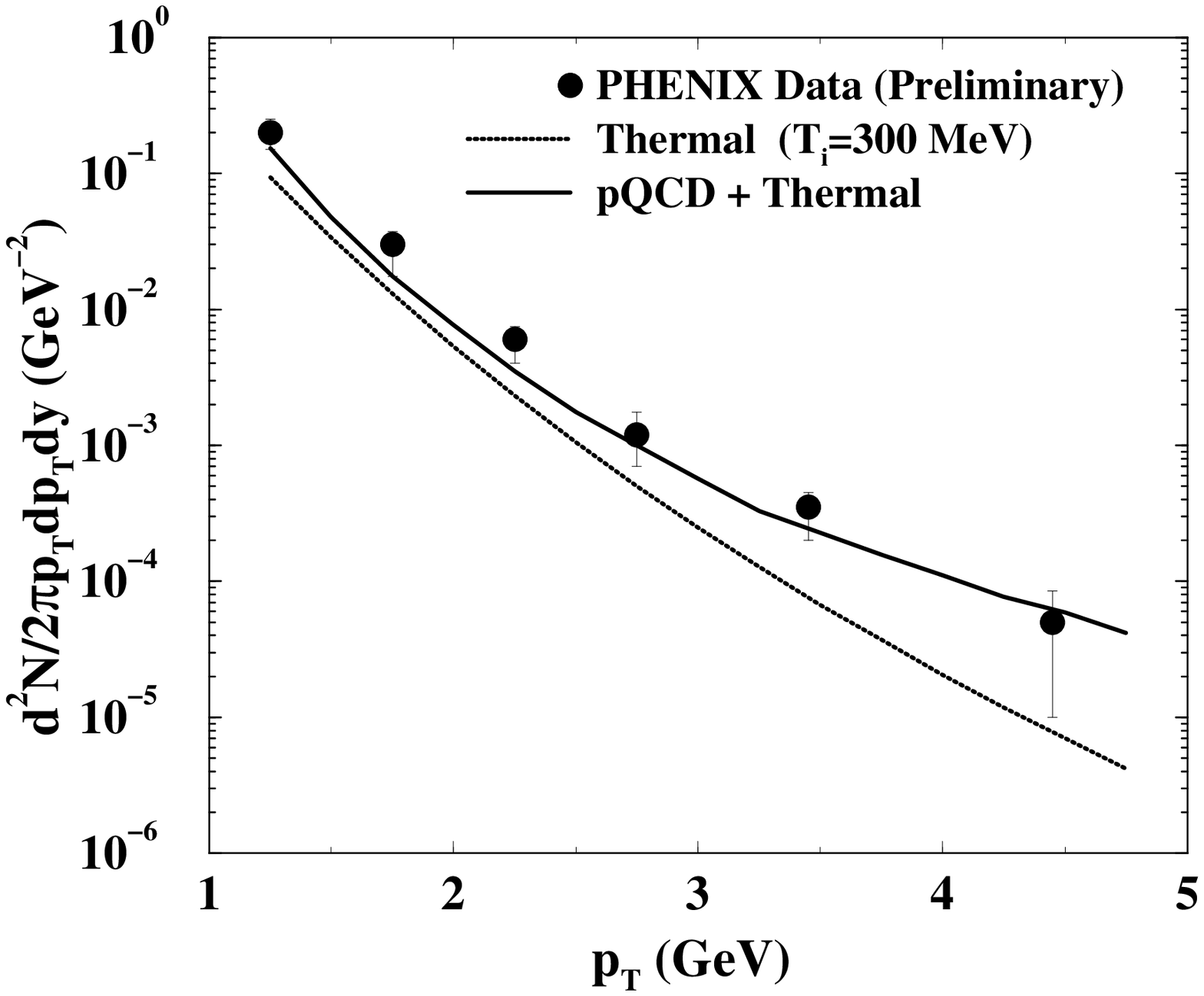}
\caption{Same as Fig.~{\protect\ref{fig2}} for type II EOS,
with $T_i=300$ MeV, $\tau_i=0.5$ fm and $T_f=120$ MeV.
Photon production rate from QGP is taken from~{\protect\cite{arnold}}.
}
\label{fig3}
\end{center}
\end{figure}
In Ref.~\cite{peress} the initial temperature and thermalization time
are taken as 590 MeV and 0.15 fm respectively to evaluate the
photon spectra. They have used the
hadronic photon production rates of Ref.~\cite{turbide}.
We reproduce the photon spectra with this initial condition 
by using the hadronic emission rates of photons
from~\cite{we1,we2,we3,we4}. As in Ref.~\cite{peress}
the medium effects are neglected in this case. 
The resulting photon
spectrum is also shown in Fig.~\ref{fig2} for comparison.
If we fix $\tau_i=0.15$ fm  then
the data can also be described reasonably 
well for $T_i=440$ MeV (keeping $T_i^3\tau_i\,\propto dN/dy$ fixed)
if the medium effects on hadrons are taken into account. 
It may be mention here that
photons from strange hadrons~\cite{turbide} is down by a factor
of 2 (at $p_T\sim 2$ GeV) compared to the production rates from 
nonstrange hadrons ($\pi,\,\rho
\,\omega,\,\eta$). The contributions involving $\eta$ mesons are neglected 
in Ref.~\cite{turbide}.

As mentioned earlier the reduction of hadronic masses in
a thermal bath increases their abundances and hence the rate of 
photon emission gets enhanced~\cite{annals,we1,we2}. 
As a result a smaller initial
temperature compared to the one obtained in Ref.~\cite{peress}, 
is seen to reproduce the data reasonably well.
The variation of hadronic masses with temperature in  
QHD model~\cite{annals, vol16} is slower than the  
BR scaling. As a result the PHENIX photon data  requires
higher value of $T_i$ in QHD  than BR scaling scenario. 
Hence to pinpoint  the actual initial temperature through photon 
spectra it is imperative to understand the properties
of hadrons in hot and dense environment. However, it is clear
that the initial temperature obtained in the present
analysis is more than the value of $T_c$ obtained from lattice QCD
calculations.  

The initial temperature obtained from the analysis 
of the RHIC data is $\sim 400$ MeV in the present work 
and $\sim$ 590 MeV in ~\cite{peress}.
This may be compared with the value of the initial 
temperature obtained from the analysis of SPS data. 
The value of initial temperature obtained  from the analysis
of the single photon data from Pb + Pb collisions at SPS~\cite{wa98} 
is within the range $\sim 200-230$ MeV ~\cite{ja1,ja2,steffen,huovinen,
KG,DYP}. A much higher value
of $T_i \sim 335$ MeV is obtained in ~\cite{dks} 
by assuming a very small value of $\tau_i\sim 0.2$ fm
for SPS energy.

To show the sensitivity of the results on the EOS we evaluate the
photon spectra using type II EOS (lattice QCD) in hydrodynamical
evolution. It is seen (Fig.~\ref{fig3}) that the data can be reproduced 
with lower
initial temperature, $T_i\sim 300$ MeV (and hence larger 
thermalization time scale $\sim$ 0.5 fm). 
This is so because in the case of type II EOS  the space time evolution 
of the system for temperature below the transition temperature is
slower than type I EOS.  Hence the hadronic phase 
live longer for type II EOS, radiating more photons from this
phase. However, it should be mentioned here that 
the slope of the $p_T$ spectra of hadrons can not be reproduced 
by type II EOS with the value of the freeze-out temperature
mentioned above.

\begin{figure}
\begin{center}
\includegraphics{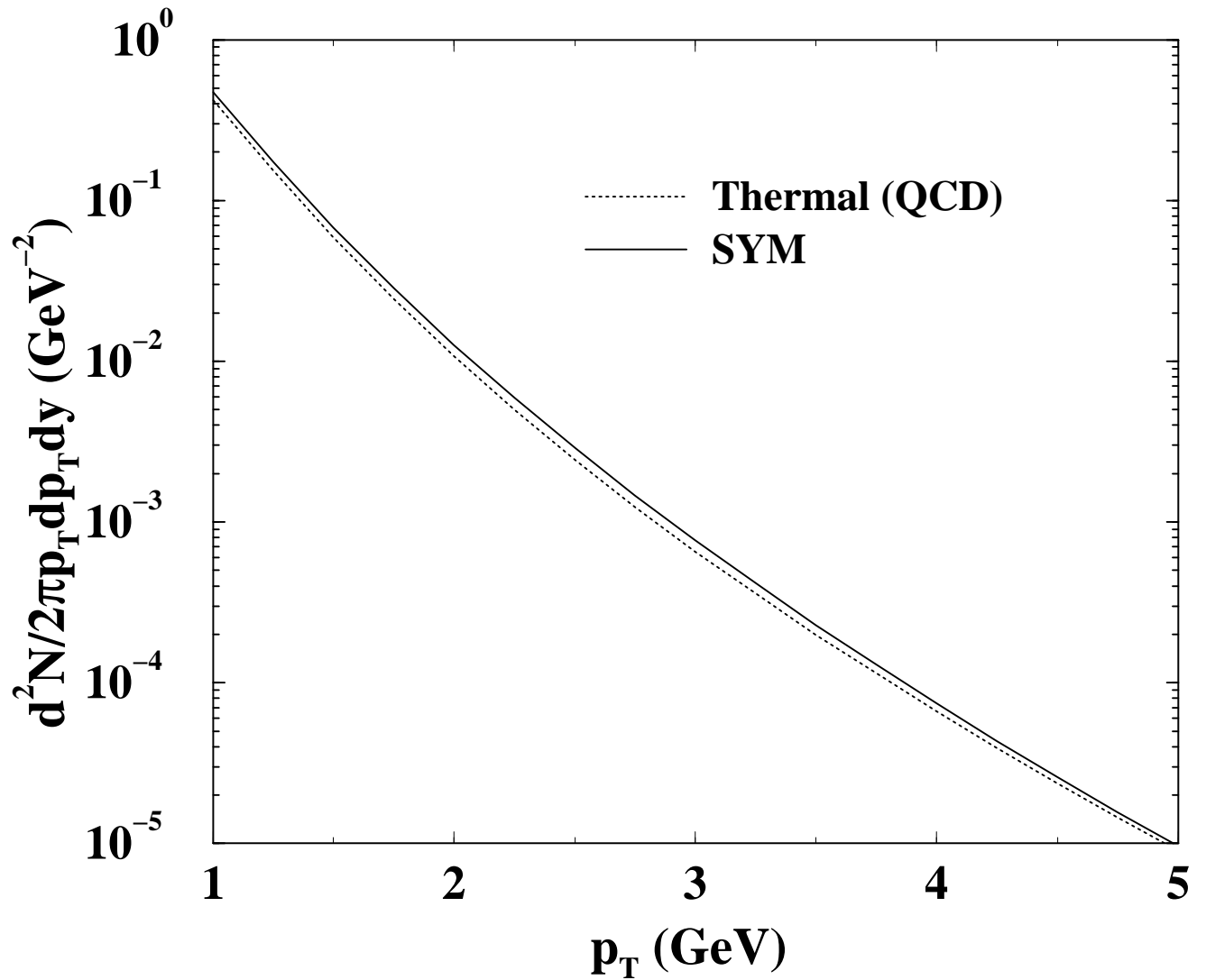}
\caption{ 
Solid (dotted) represents  $p_T$ spectra of thermal photons
when photons from QGP is evaluated within the ambit of  SYM theory 
(thermal QCD).  The values of $T_i=590$ MeV,  $\tau_i=0.15$ fm
and $T_f=120$ MeV. Type II EOS has been used here.
}
\label{fig4}
\end{center}
\end{figure}

It may be mentioned at this point that the photon emission
rates obtained in~\cite{arnold}  are valid in weak coupling
limit, although the QGP formed after $Au + Au$ collisions
at RHIC energy could be strongly coupled\cite{strong}. However, photon
production from strongly coupled QGP is not available from thermal
QCD. Therefore, results in strong coupling limit would be
useful even if it comes from a theory which is not real QCD.
Recently, results from ${\cal{N}}=4$ Supersymmetric Yang Mills (SYM) theory
have been made available~\cite{kovtun,huot} in the strong coupling limit. 
The rate obtained in this case could be treated as a
upper limit of photon production from QGP.  The thermal photons obtained
in this case are compared with that from thermal  QCD in Fig.~\ref{fig4}. 
Photons from SYM is enhanced by about $20\%$ as compared to thermal
QCD in the $p_T$ region $\sim 2 $ GeV.

\begin{figure}
\begin{center}
\includegraphics{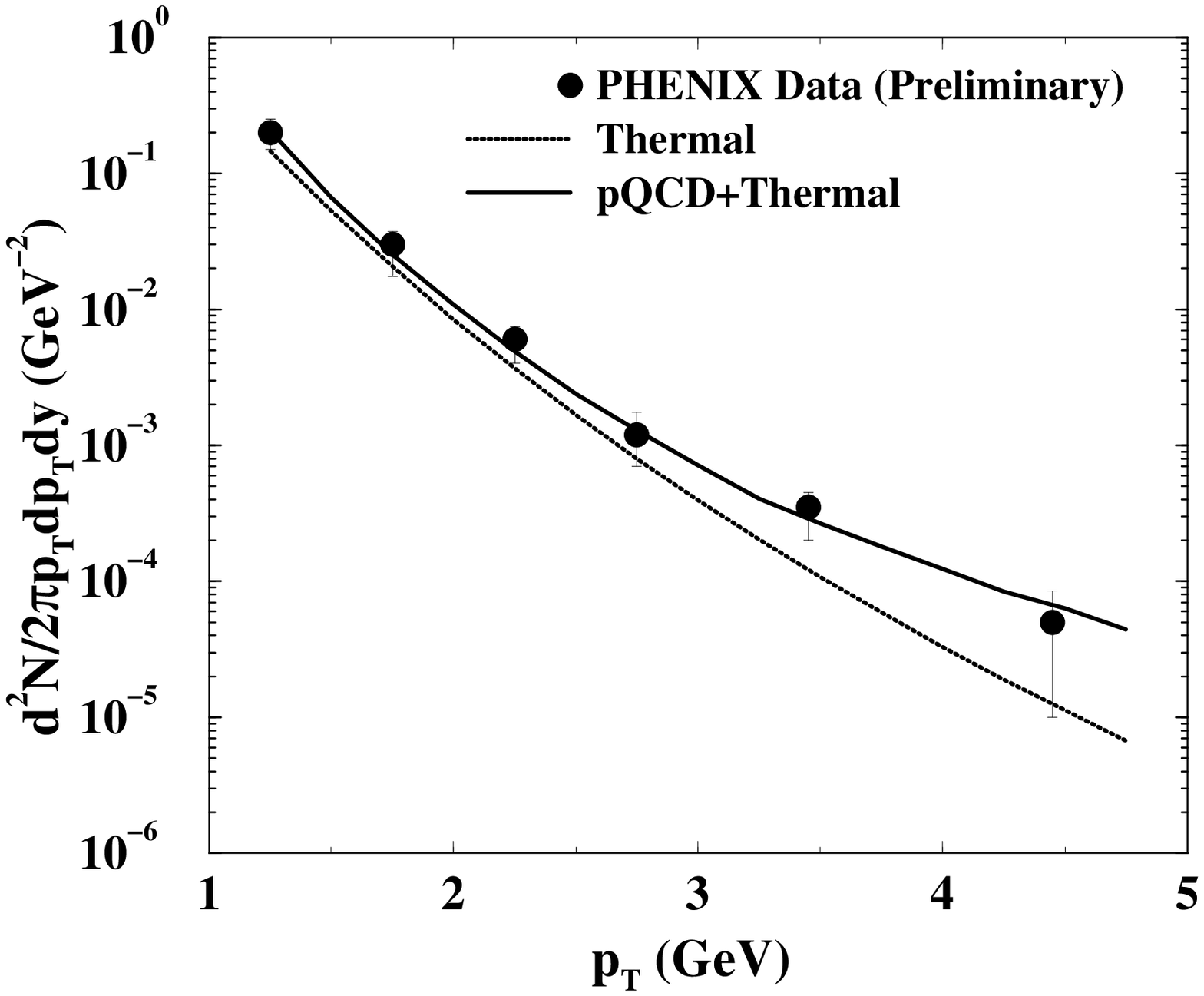}
\caption{Direct photon spectra at RHIC energies measured by 
PHENIX Collaboration.
Solid (dotted) line depicts the pQCD + thermal (thermal) photon yield. 
Thermal photons from QGP phase  is obtained from SYM theory~{\protect\cite{arnold}}.
Here $T_i$ = 300 MeV and
$\tau_i=0.5$ fm. Type II EOS is used to obtain the thermal contributions. 
}
\label{fig5}
\end{center}
\end{figure}

When the results obtained from SYM is added with thermal photons from
hadronic phase and photons from pQCD it describes the data reasonably 
well (see Fig.~\ref{fig5}). The initial temperature and time are taken 
as 300 MeV and  $0.5$ fm respectively. The type II EOS is used here.
It should be mentioned here that 
for type I EOS and production rate from SYM, 
$T_i\sim 400$ MeV and $\tau_i\sim 0.2$ 
fm is required to reproduce the data.

\begin{figure}
\begin{center}
\includegraphics{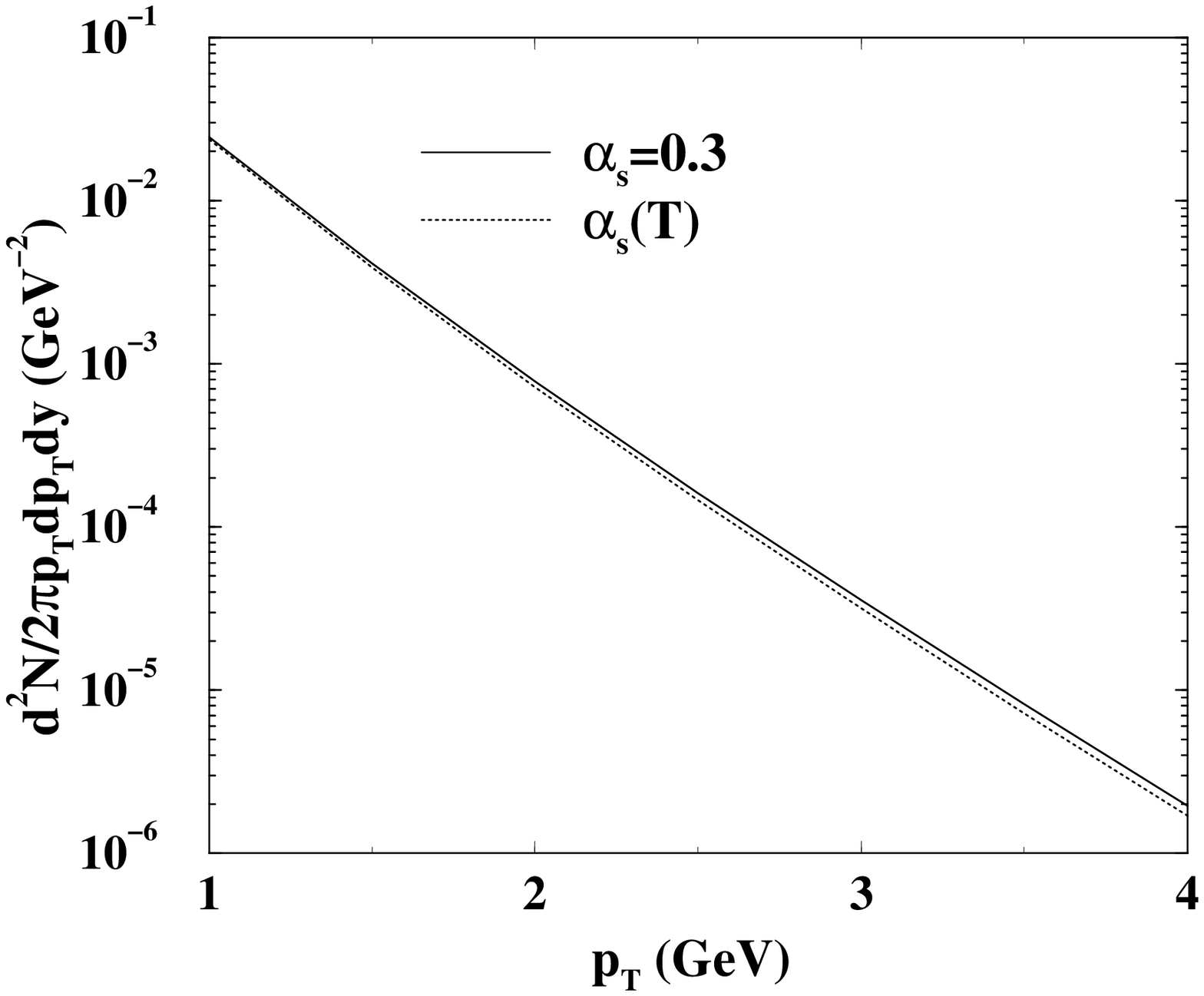}
\caption{Photon emission from QGP phase for two values of 
strong coupling constant $\alpha_s$. Solid (dotted)
line indicates results for $\alpha_s=0.3$ (temperature dependent 
coupling). Here $T_i=400$ MeV, $\tau_i=0.2$ fm and $T_f=120$ MeV.
Type I EOS has been used here.
}
\label{fig6}
\end{center}
\end{figure}

In Fig.~\ref{fig6} the dependence of photon spectra from QGP phase 
on strong coupling is demonstrated. The temperature dependence 
of $\alpha_s$ has been taken from ~\cite{zantow}. The difference
in photon spectra at $p_T\sim 3$ GeV for $\alpha_s=0.3$ and
temperature dependent $\alpha_s$ is about $13\%$. 

\begin{figure}
\begin{center}
\includegraphics{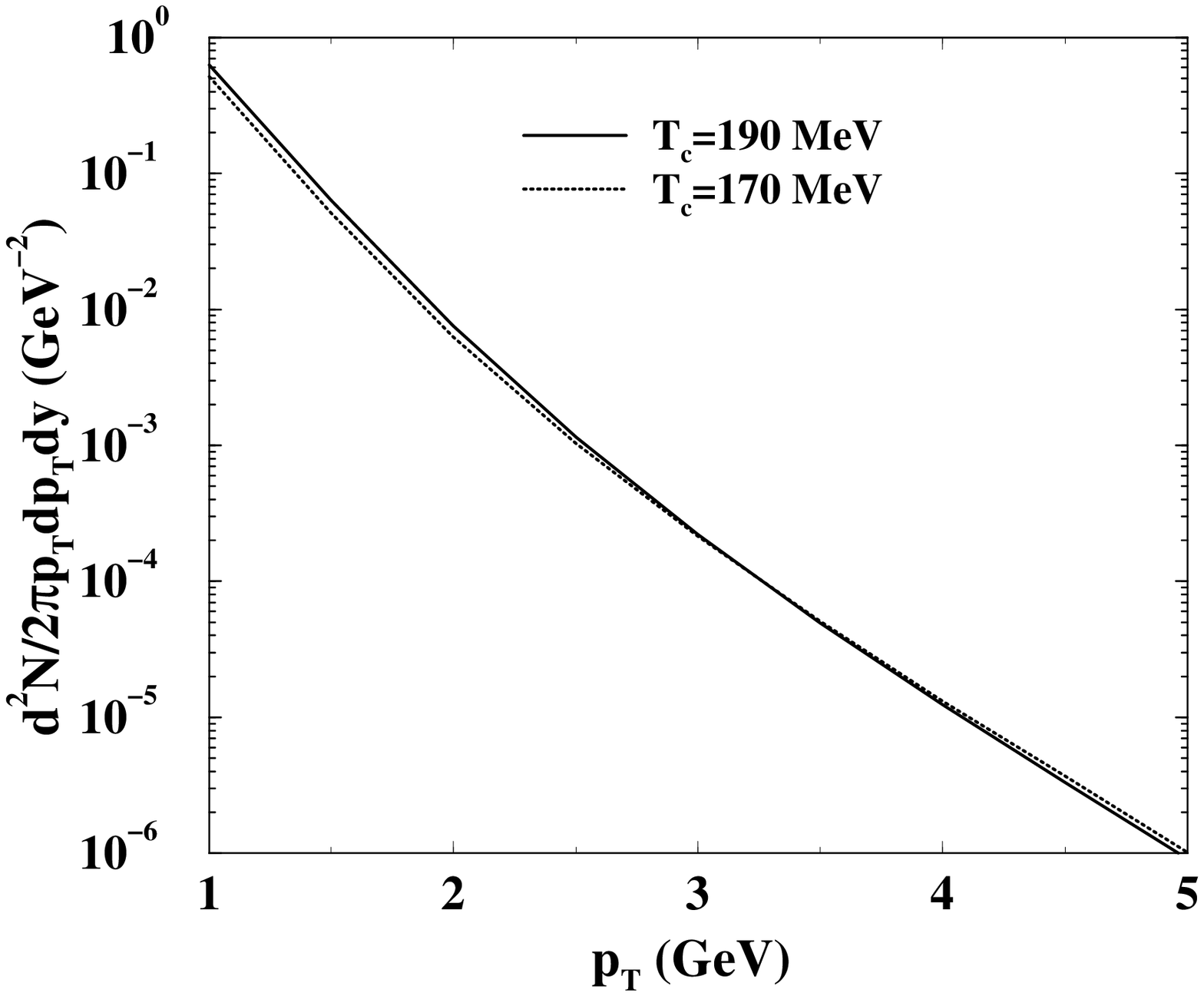}
\caption{Photon emission from QGP phase for two values of 
transition temperature, $T_c$. Solid (dotted)
line indicates results for $T_c=190 (170)$ MeV.
Here $T_i=400$ MeV, $\tau_i=0.2$ fm and $T_f=120$ MeV.
Type I EOS has been used here.
}
\label{fig7}
\end{center}
\end{figure}

A $20\%$ enhancement is obtained at $p_T\sim 2$ GeV in total thermal photon 
production if transition temperature is increased from 170
to 190 MeV (Fig.~\ref{fig7}). Photons from hadronic phase 
populate mainly the low $p_T$ region of the spectra. 
Larger value of transition temperature  means that 
hadrons survive up to larger temperature  
and emit more photons at low $p_T$ region.
 
\begin{figure}
\begin{center}
\includegraphics{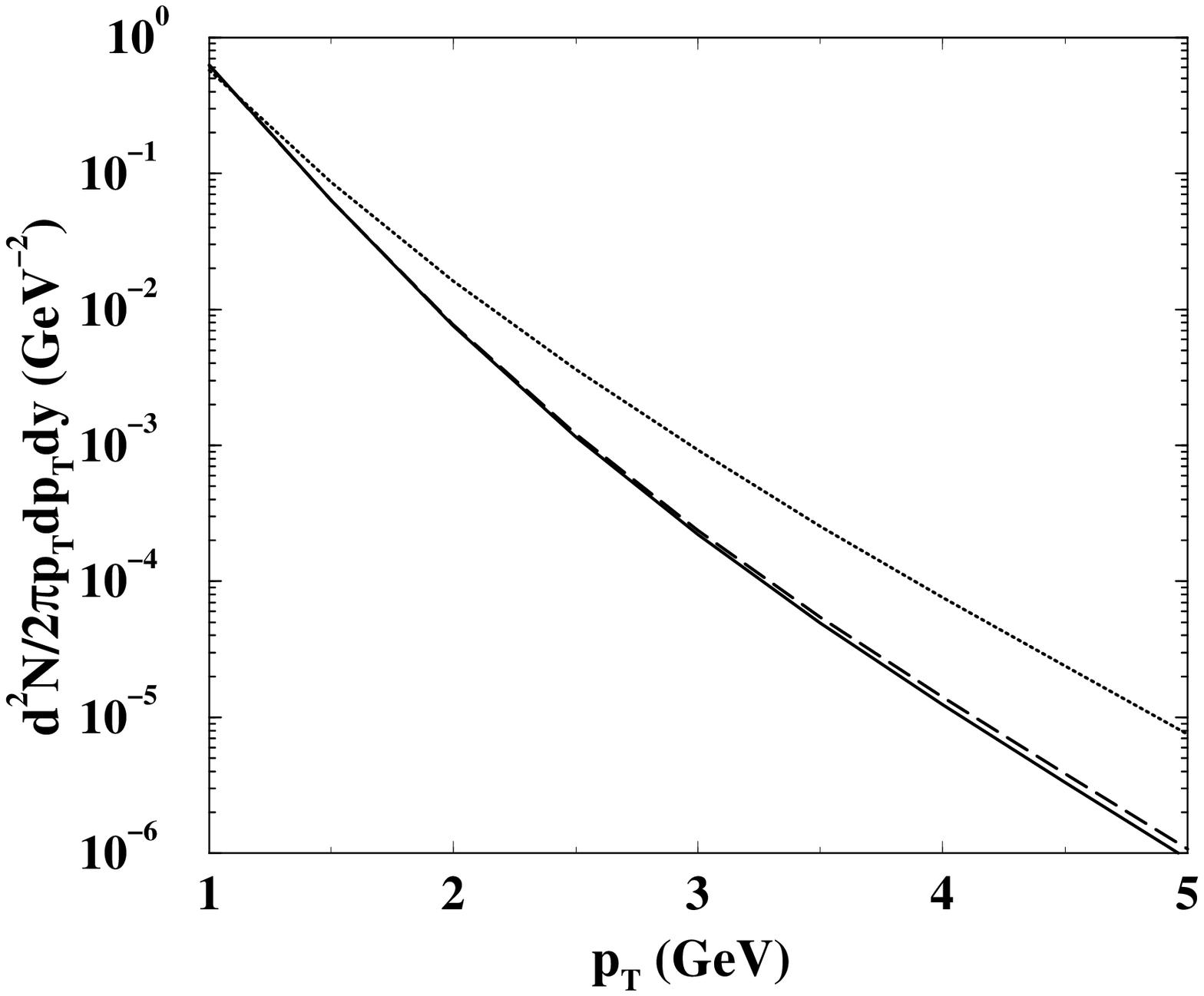}
\caption{Thermal photon spectra with different 
initial velocity profile. Solid (dashed) line indicates results for
$v_0=0 (0.2)$ in~Eq.\protect\ref{vrin}. 
Results for a different initial velocity profile
$v_r(\tau_i,r)=v_0^1\,r/R_A$ with $v_0^1=0.2$ is also shown (dotted line).  
Here $T_i=400$ MeV, $\tau_i=0.2$ fm and $T_f=120$ MeV.
Type I EOS has been used here.
}
\label{fig8}
\end{center}
\end{figure}

All the results presented above are obtained with vanishing
initial radial velocity {\it i.e.} for $v_0=0$ in Eq.~\ref{vrin}.
Finally, we demonstrate the sensitivity of results on the value 
and shape of the initial velocity profile. In Fig.~\ref{fig8}
we show the results for $v_0= 0$ (solid line)
and $v_0=0.2$ (dashed line) in Eq.~\ref{vrin}. The difference in results is
rather small. However, for a different velocity profile, $v_r(\tau_i,r)=v_0^1\,r/R_A$
a substantial change in the spectra is observed for $v_0^1=0.2$, because
this gives a stronger radial velocity 
distribution of the fluid compared to Eq.~\ref{vrin}. 
It will be interesting to put constrains on the initial velocity
profile from experimental data on the $p_T$ distributions of
various types of hadrons~\cite{lpairs}.

\section{Summary and Discussions}
In summary, we have analyzed the direct photon data measured
by PHENIX collaboration for $Au + Au$ collisions at $\sqrt{s_{NN}}=200$ GeV.
The data can be reproduced by assuming a deconfined state of quarks
and gluons with initial temperature $\sim 400$ MeV.
However, for type II EOS the data can be explained for
lower value of $T_i\sim 300$ MeV. 
Since the type II EOS fails to reproduce the slope of
the $p_T$ distributions of pions and kaons as mentioned 
earlier, this value of
$T_i$ should be taken with caution. 
The type I EOS considered here
is similar to EOS Q of ref.~\cite{huovinennpa}. EOS H of
~\cite{huovinennpa} (hadronic
scenario without any phase transition) has not been
considered here, because we feel this scenario is not
realistic at RHIC collision at $\sqrt{s_{NN}}=200$ GeV. The other
EOS, thermal quasi-particle model 
considered in~\cite{schneider} gives $T_i$ more than 300 MeV. Therefore,
$T_i=300$ MeV obtained here for type II EOS is the conservative lower 
limit.  Photon productions from thermal QCD
and ${\cal{N}}=4$ SYM have  been compared to the data.
In both the cases similar values of the initial temperatures are
obtained.
The extracted average temperature ($T_{av}$)
from the slope of the photon $(p_T)$ spectra is found to 
be $\sim$ 265 MeV for the $p_T$ range between 1.25 to 2.25 GeV
where thermal contributions dominate. When the effects of flow
is ignored (by putting $v_r=0$ for all time)  the `true' average 
temperature is found to be 215 MeV. 
The initial temperature must be more than 215 MeV
because the thermal photon spectra
is a superposition of emission rates for all
the temperatures from initial to freeze-out. 
This indicates that
the temperature of the system formed after the collisions
is higher than the transition temperature for deconfinement. 
The lower limit in
the initial temperature can be used to put a conservative
upper bound on the thermalization time, which in the 
present case is $\sim 1.3$ fm.
The invariant mass distribution of lepton pairs in the similar
framework will be reported shortly~\cite{lpairs}.

In spite of the encouraging situation described above
it is worthwhile to mention the following.
The experimental data~\cite{phenix} for real photon spectrum in $Au+Au$ collision
has been obtained from the analysis of the low mass and high $p_T$ lepton
pairs by assuming $\gamma_{direct}/\gamma_{incl}
=\gamma^*_{direct}/\gamma^*_{incl}$ ($*$ indicates virtuality). 
The similar procedure should
be adopted for analysis of photon data for $pp$ and $p(d)+Au$ collisions.
Moreover, the data from $pp$ and $p(d)+Au$ collisions for $1<p_T$(GeV)$<5$ 
at $\sqrt{s_{NN}}=200$ GeV will be useful to validate  the NLO pQCD 
contributions for $Au+Au$ collisions and hence will be helpful
to quantify the thermal contribution.

{\bf Acknowledgment:} One of us (JA) is grateful to 
Pavel Kovtun for fruitful discussions on photon production in
supersymmetric Yang-Mills plasma. JA would also like to
thank nuclear theory group, Brookhaven National Laboratory
for their kind hospitality during his stay 
where part of this work was done.

\end{document}